# Efficient Traffic Control of VoD System

Soumen kanrar

Vehere Interactive Pvt Ltd, Calcutta-53, India
`Soumen.kanrar@veheretech.com`

## ABSTRACT

*It has been a challenging issue to provide digital quality multimedia data stream to the remote user through the distributed system. The main aspects to design the real distributed system, which reduce the cost of the network by means of reduce packet loss and enhanced over all system performance. Since the number of user increased rapidly in the network it posed heavy load to the video servers. The requested clients, servers are all distributed in nature and the data stream delivered to the user without error. In this work I have presented the performance of the video on demand server by efficient traffic control at real time with respect to incoming multirate traffic pattern . In this work, I present how the overall system performance gradually decreases when the client population sized in the clusters increase. This work indicated the load balancing required for the on demand video distributed system to provide efficient cost effective service to the local or remote clients.*

## 1. INTRODUCTION

Analyzing and performance measurement are the main focused area of distributed video on demand system. The distributed video on demand (VoD) system becomes important services supported by the high-speed networks, video servers and distributed multimedia file systems. A client will be able to request a video from any where and at any time. In response to a client's request, VoD systems will deliver high quality digitized video directly to client set-top-box. The distributed VoD architecture consists of cluster of clients, networks and cluster of video servers. The set-top-boxes at the client side providing buffer for periodically delivered video segments from the video servers. The video stream are collected partly or completely by the local cluster .The service from the local cluster is provided over a local distributed network, such as an ATM, LAN or xDSL, HFC. It is assumed that there are sufficient network resources at the local distributed network to deliver video to the clients and that there is no resource contention on the local distributed network. This is a reasonable assumption, because the local cluster acts as a neighborhood cluster of servers and the overall VoD user population grows more new local clusters are added. When a new neighborhood of local cluster is added to the distributed VoD system, the network capacities can be sized to match the reference user population for that neighborhood. Clearly it indicate the scalability of the distributed on demand video system. The requests originating from a reference user population are best served by its own local sites, because of the absence of network contention. The clusters of servers administer admission control tests before accepting new requests. The remote site may be archival in nature, providing a permanent repository for all videos or they may act as replicated servers such as mirrored sites. Remote servers also provide video delivery service over the high speed networks connecting to remote sites. The remote sites may provide service to many user populations. The distributed system itself is a hierarchy of neighborhoods in the geographical region. The hierarchy of clusters of servers, cluster of user population and connection networks is scalable. If a request cannot be served from the local site, it may be directed to other remote site. Each user request is assigned its own video data stream from either the local or the remote server. A typical Video on Demand (VoD) service allows remote users to play back any one a large collection of videos at any time. Typically, these videos files are stored in a set of central video serves and distributed through high-speed communication networks to geographically – dispersed clients. The server sends the sequential video stream packets after received the client request. Each video stream can be viewed as a concatenation of a storage –I/O "pipe" and network pipe. So sufficient bandwidth and storage space are required at the network interface card (NIC) for received the video stream, storing the forwarding that sequentially. Thus, a video server has to reserve sufficient I/O and network bandwidths before accepting a client's request. So a dedicated server channel required for continuous flow of the sequential video stream that can be played back continuously at the client side. The video on demand system mostly applied the service sector like medical information service, distance learning, home entertainment, digital video library, movie-on-demand, Tele-shopping, news-on-demand. In general, the VoD service can be characterized as long lived session of High bandwidth requirements and quality sensitive service. In long lived session, a VoD system should support long lived session. For example, a typical movie on-demand service runs for 1 hour to 2 hours. For the High bandwidth requirements, server storage I/O and network bandwidth requirements more than one mega bits per seconds. A client requires the VoD system to offer VCR –like interactively such as ability to play, forward, reverse and pause. Other advanced interactive feature includes the ability to skip or select advertisements. This is related to the quality sensitive service. The quality of service that VoD consumers and service providers might care includes service latency, defection rate, interactivity, playback effects of videos. In order to support a large population of clients, we therefore need solution that efficiently utilizes the server and network resources. Class based admission control [1] used to the video on demand system to get better performance. Some planning and managing of the resource in the Internet based [3 ,4] Video on demand give better performance. In fact, the network I/O bottleneck has been observed in many earlier systems, such as Network project in Orlando [ 7] and Microsoft's Tiger Video Fileserver [8]. In order to support a large population of clients, we therefore need new solutions that efficiently utilize the server and network resources. The multicast facilities of modern communication network [6,9,10,12] offers an efficient means of one to many data transmission. The enhancement of the system performance depends upon the reduction of the retransmitted packets through the links passing through the routers. Multicast can significantly improve the VoD performance by reduces the required network bandwidth greatly, so the overall network load be reduce. In other way the multicasting alleviates the workload of the VoD server and improves the system throughput by batching requests. Multicasting offer excellent scalability which in turn, enables serving a large number of clients that provide excellent cost -performance benefits. In this paper I have shown how the traffic load to the server can be control by used of some control based on Traffic handle. This work presents significant impact of the Traffic handle on the performance of the VoD system. This paper is structured as follows. Section 2 presents the basic Network architecture for the video on demand distributed system. In this section, briefly represent the Hierarchical architecture of the system and illustrate the analytic from of the request control of the video server. Section 3 presents the parameters description of the simulation environment. Section 4 presents the simulation results with respect to the traffic control in the distributed video on demand network architecture. Section 5 presents conclusion remarks and references.

## 2. The architecture of Distributed VoD system

Large-scale VoD system requires the servers to be arranged as a distributed system in order to support a large number of concurrent streams. The system is hierarchical and an end node server handles the requests from particular area. The next server in the hierarchy takes the requests over for end node servers if they cannot handle them. This architecture provides the cost efficiency, reliability and scalability of servers. Generally, servers are either tree based shaped [16] or graph –structured [11,14, 15] The graph –structured system often offer good Quality of Service for handling the requests. But the management of request, videos and streams is complicated in the system. The tree-shaped system can easily manage requests, videos and streams, but it offers poorer quality of service than the former. In order to evaluate the effectiveness of distribution strategies in such a hierarchy, the author of [16] investigate how to reduce storage and network costs while talking the customer's behaviors into account.

### 2.1. Hierarchical VoD Architecture Model

Although some of hierarchical architectures are originally designed for unicast VoD services, they can also be used for multicast VoD to further improve the efficiency of service.

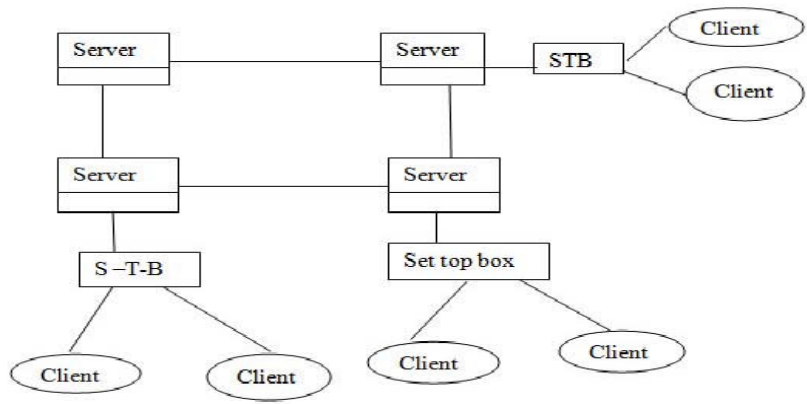

Figure- 1

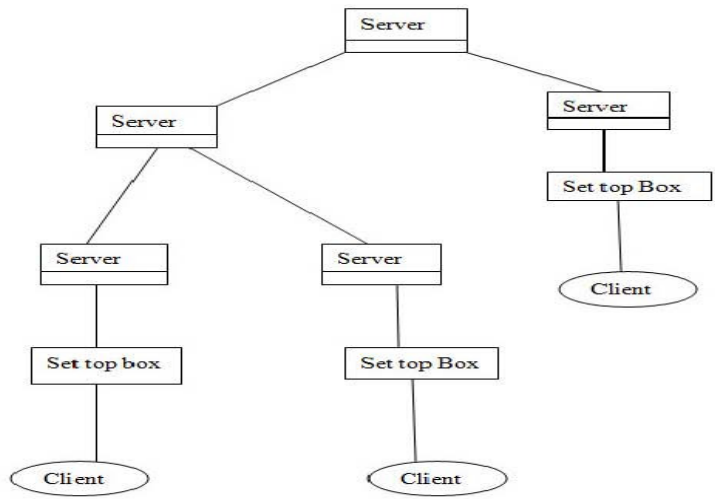

Figure - 2

The proposed model of the Distributed video on demand architecture satisfies the basic requirements of the distributed networks. The distributed system impose minimal additional burden on the server and the networks. The system sufficiently utilizes critical resources on the server and the network. The distributed system responded to the consumer requests and transmits the requested videos in real time. The distributed system can scale well with the number of clients. The system provides the clients full control of the requested video by using interactive functions. The most important part of the distributed system is the reliability. The system easily recovers from failures and the transmission of messages and video streams should also be reliable.

## 2.2. Analytic Model

Let the maximum number of requests, $n$ serviceable at the server. If $R_d$ be the overall disk bandwidth and $R_p$ is the client request play back rate then, $n \leq \left\lfloor \dfrac{R_d}{R_p} \right\rfloor$ [1] and admission control test at the server to determine whether the server able to serve maximum n request. We proposed that a VoD server which has a total bandwidth capacity $C$ ports. Let us have $k$ classes of service requests for video streams each with rate $\lambda_i$ where $i$=1 to $k$. The servers capacity is divided into sections with each section has ports capacity $C_j$ where j=1 to $k$. Via-figure 3. Total system capacity, $C = \sum_{j=1}^{k} C_j$

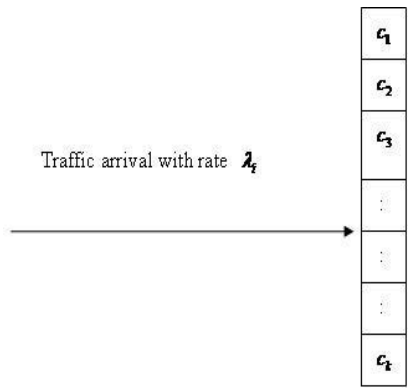

Traffic Arrival, Figure 3

Request from the class a will be admitted to a partition b with probability
$$p^{\#}(a,b) = p_a^{\#} \qquad (1)$$

Where, $\sum_{i=1}^{k} p^{\#}_i = 1$, A new request arrives with Poisson distribution for video stream. We assume the server ports occupancy for class $i$ request is $Q_i$. When a request of class $i$ arrives, we check whether $Q_i \prec C_i$ if so, then it is admitted with probability $p_a^{\#}$. If request is served and then $Q_i = Q_i + 1$. If $Q_i \geq C_i$ then we check if $Q_{i+1} \prec C_{i+1}$, Then the request is admitted by the server with probability $p^{\#}_{a+1}$ and $Q_{i+1} = Q_{i+1} + 1$ otherwise if the request is not admit, repeat for $Q_{i+2}$. If the request can't be admitted by all the sections, then the request consider to be blocked and the server will discard the request (via figure 4)

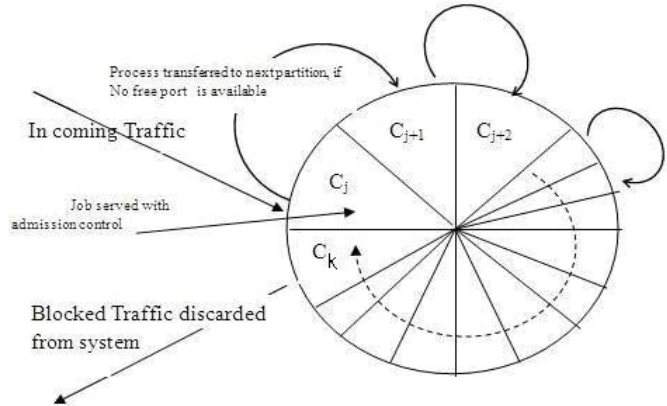

Process State in Server Figure 4

Let request comes from $i^{th}$ class of population and served by the $j^{th}$ partition block of the server clearly $1 < j$. Where $A_j B_i B_{i+1}............B_{j-1}$ is the event that the previous all partition from $i$ to $j-1$ is blocked only $j^{th}$ partition has at least one available port. [2]

$B_i B_{i+1}.......B_{j-1}$ represents the event that all the partition from $i$ to $j-1$ is blocked. Now by considering the expression (1) we get,

$$p(A_j / B_i B_{i+1}............B_{j-1})$$
$$= (p^{\#}_a) \frac{p(A_j B_i B_{i+1}.........B_{j-1})}{p(B_i B_{i+1}........B_{j-1})}$$
$$= (p^{\#}_a) \frac{p(A_j) p(B_i) p(B_{i+1})........p(B_{j-1})}{p(B_i)..............p(B_{j-1})}$$
$$= (p^{\#}_a) p(A_j) \qquad [$$

Where, a = 1…...n and 1<= ( i , j) <= n.

Now for the over all distributed system, if $R_r$ is reserved bandwidth rate can be expressed as

$$R_r = \frac{KB_r + (J-K)M}{\gamma} \quad (2)$$

Where $J$ is the total number of links, $K$ is the number of link used for client in play back or interactive session. $\gamma$ be the play back duration for play back or interactive session. $B_r$ is the maximum amount of data sent to any interactive session. The reserve bandwidth can't be grater than the overall bandwidth of each link on the paths with J links. Let $A_j$ is the overall bandwidth on the $j^{th}$ links clearly $R_r \leq A_j$ for $j \in [1,..J]$

## 3. Explanation of simulation environment

The simulation has done in two parts. In the first part, the randomly generated traffic handles probabilities on the basis arrival traffic rate. In the second part holds the simulation parameters for the on demand video servers. According to the figure 2, I have consider, the k number of sectors and each sector holds n number of ports. The first part of the simulation generated $k \times n$ size of control matrix. Table 1 presents control probabilities for k=20 and n =10.

Table 1. Control Matrix

| 0.04 | 0.04 | 0.03 | 0.03 | 0.04 | 0.04 | 0.04 | 0.03 | 0.05 | 0.07 |
|---|---|---|---|---|---|---|---|---|---|
| 0.04 | 0.04 | 0.02 | 0.03 | 0.08 | 0.03 | 0.04 | 0.06 | 0.06 | 0.01 |
| 0.09 | 0.08 | 0.1 | 0.04 | 0.09 | 0.01 | 0.07 | 0.04 | 0.07 | 0.09 |
| 0.04 | 0.01 | 0.04 | 0.04 | 0.04 | 0.04 | 0.08 | 0.07 | 0.03 | 0.03 |
| 0.06 | 0.01 | 0.07 | 0.07 | 0.09 | 0.06 | 0.08 | 0.06 | 0.04 | 0.08 |
| 0.02 | 0.06 | 0.05 | 0.05 | 0.07 | 0.06 | 0.01 | 0.05 | 0.05 | 0.06 |
| 0.08 | 0.01 | 0.1 | 0.07 | 0.08 | 0.08 | 0.06 | 0.07 | 0.01 | 0.06 |
| 0.04 | 0.06 | 0.07 | 0.02 | 0.09 | 0.05 | 0.02 | 0.07 | 0.06 | 0.06 |
| 0.05 | 0.01 | 0.03 | 0.05 | 0.06 | 0.08 | 0.08 | 0.06 | 0.04 | 0.07 |
| 0.04 | 0.05 | 0.03 | 0.03 | 0.06 | 0.08 | 0.05 | 0.05 | 0.02 | 0.02 |
| 0.04 | 0.02 | 0.06 | 0.04 | 0.03 | 0.07 | 0.06 | 0.04 | 0.05 | 0.05 |
| 0.05 | 0.06 | 0.07 | 0.1 | 0.04 | 0.08 | 0.02 | 0.02 | 0.04 | 0.02 |
| 0.1 | 0.08 | 0.04 | 0.03 | 0.08 | 0.08 | 0.04 | 0.06 | 0.06 | 0.07 |
| 0.04 | 0.07 | 0.05 | 0.04 | 0.01 | 0.05 | 0.08 | 0.08 | 0.02 | 0.04 |
| 0.09 | 0.06 | 0.02 | 0.1 | 0.06 | 0.02 | 0.07 | 0.03 | 0.08 | 0.07 |
| 0.02 | 0.08 | 0.03 | 0.09 | 0.03 | 0.03 | 0.08 | 0.02 | 0.07 | 0.01 |
| 0.06 | 0.09 | 0.07 | 0.02 | 0.02 | 0.08 | 0.04 | 0.04 | 0.04 | 0.01 |
| 0.06 | 0.04 | 0.04 | 0.03 | 0.01 | 0.01 | 0.03 | 0.02 | 0.07 | 0.08 |
| 0.03 | 0.08 | 0.03 | 0.02 | 0.01 | 0.02 | 0.02 | 0.08 | 0.08 | 0.08 |
| 0.01 | 0.07 | 0.05 | 0.09 | 0.01 | 0.03 | 0.03 | 0.05 | 0.06 | 0.02 |

In the second part of the simulation i have presented the simulation parameters for the video on demand server. For the simplicity, I have considered that each sector of the server contained equal number of ports. The traffic arrived rate from different cluster of clients, started at 1 Mb/s and end at 10.5 Mb/s. The number of clusters of client is 20. Table 2 presents the overall simulation parameters for the VoD server.

International Journal of Computer Networks & Communications (IJCNC) Vol.3, No.5, Sep 2011

Table 2. Parameters of simulation environment

| Number of Clusters | 20 |
|---|---|
| Traffic Arrival Rate (Minimum) | 1 Mb/sec |
| Traffic Arrival Rate (Maximum) | 10.5 Mb/sec |
| Number of ports in each section | 10 |
| Port access time | 140 sec |
| Number of section | 20 |
| Simulation Time | 460 sec |

## 4. Performance Analysis

Figure 5 to 9 represents the performance of distributed video on demand system with respect to the efficient control of traffic in the network. Figure 1 represents the video on demand system with respect to the incoming traffic flow from the 20 cluster of clients. The rate of the traffic started 1Mb/second from the first cluster. 1.5 Mb/second from the second cluster and 10.5 Mb/second from the $20^{th}$ client cluster.

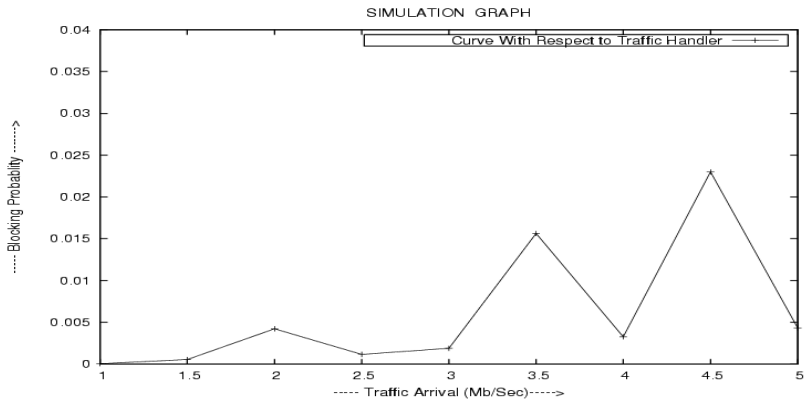

Figure 5

The figure 5 represent the performance of the distributed video on demand system with respect to the control probability set {0.04, 0.04, 0.08,….0.07} selected from the $2^{nd}$ column of the table 1.

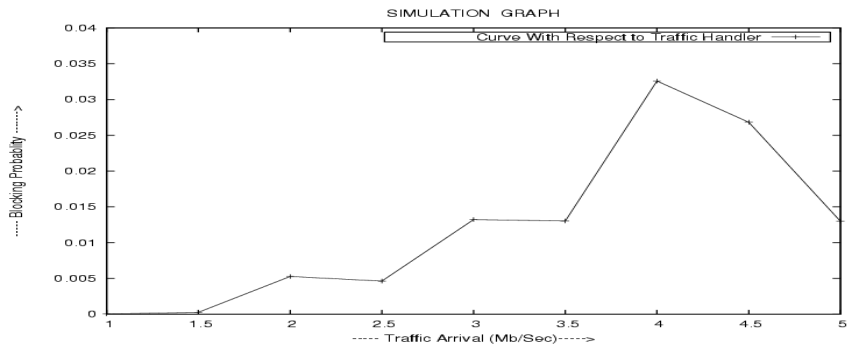

Figure 6

Figure 6 presents the performance of the distributed video on demand system with respect to the control probability set {0.03, 0.02, 0.1, 0.04….0.03, 0..05} . The traffic arrival rate to the video system 1 Mb/second 1.5 Mb/second …. 10.5 Mb/second.

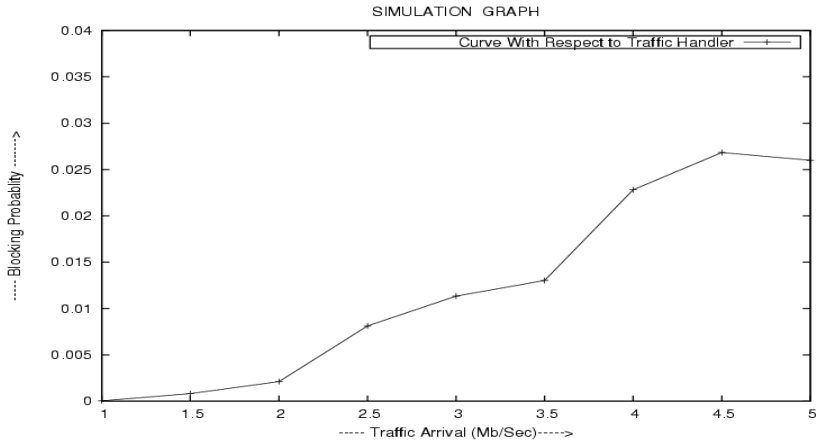

Figure 7

Figure 7 presents the performance of the distributed video on demand system with respect to the control probability set {0.3, 0.06, 0.04,...,0.08,.0.05} . The traffic arrival rate to the video system 1 Mb/second 1.5 Mb/second up to 10.5 Mb/second.

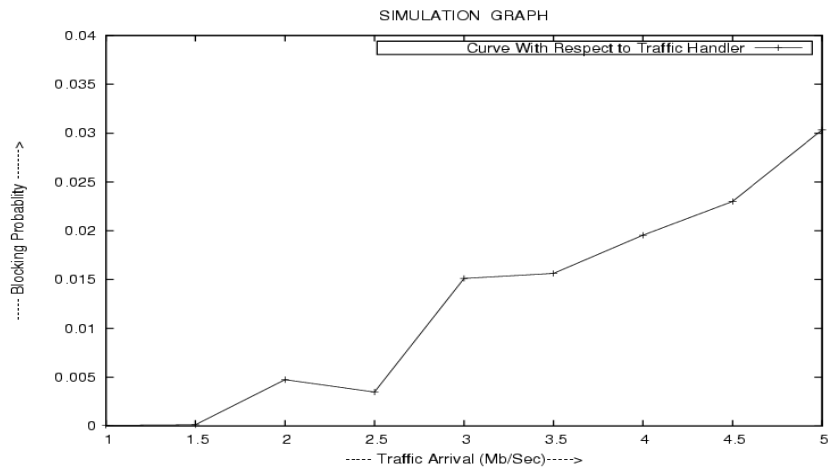

Figure 8

Figure 8 presents the performance of the distributed video on demand system with respect to the control probability set {0.07, 0.01, 0.09… 0.08, 0.02}. The traffic arrival rate to the network system starts from 1 Mb/second 1.5 Mb/second up to 10.5 Mb/second. Figure 9 and figure 10 represents the comparison study of the performance of the video on demand system with respect to the policy based traffic handling and without policy based traffic handling.

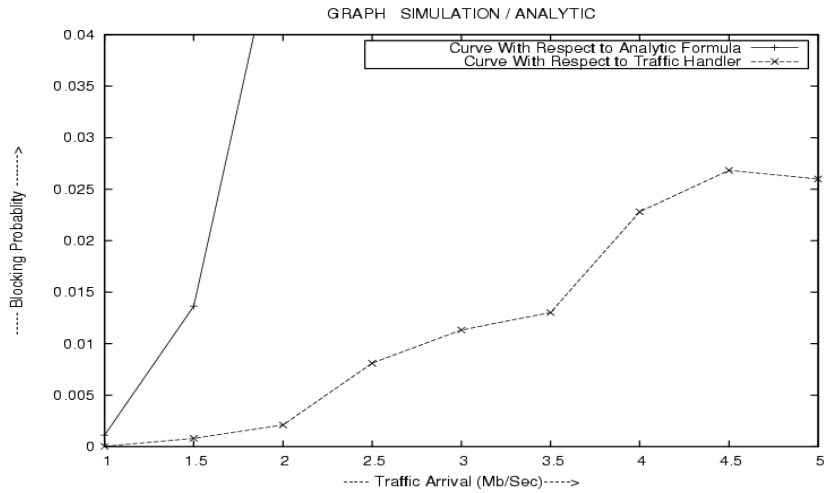

Figure 9

Figure 9 and figure 10 represents two types of graph, when the policy based traffic handling not used the blocking on the video on demand system increase very rapidly. From the figure 5 and figure 6, we see the blocking curve presents like a exponential growth. When we used policy based traffic handle, the blocking curve is below some threshold level. Clearly the request can be handling very efficiently by using this control mechanism. On the other hand the load on the video on demand system decrees. Effectively it will increase the overall system performance. The packet loss reduced efficiently by the policy based traffic control.

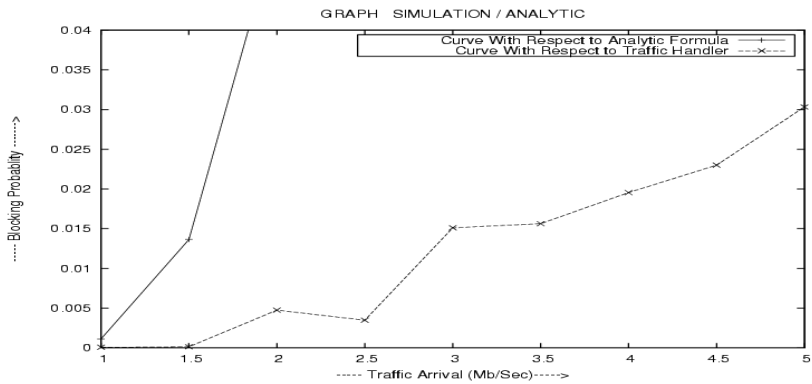

Figure 10

Figure 11 represents the comparison study of the system throughput of two types of client clusters. The 1st strip show the corresponding throughput of the system for the smaller size of population cluster with respect to the request submitted and the request accept by the system. The 2nd strip represents the system throughput of the system, when the cluster of population grater then the previous one. The simulation result clearly indicated if the size of the population in the cluster increases the system performance gradually decreases.

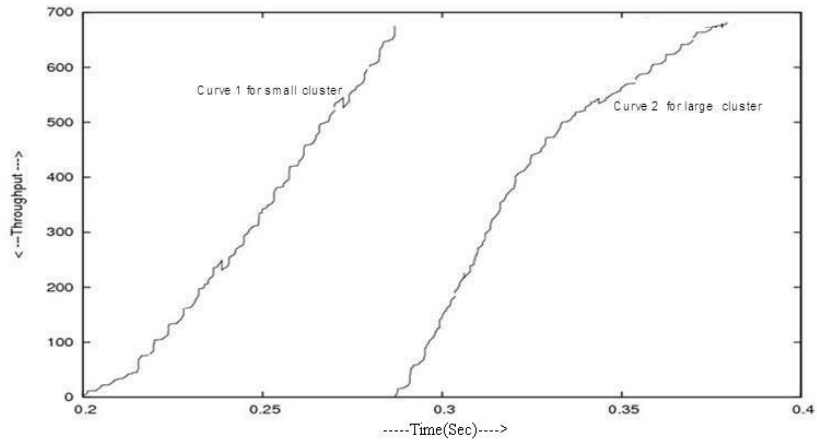

Figure 11

Figure 12 represents the traffic intensity i.e. the traffic load inside the VoD system with respect to the incoming traffic. The figure 8 represents the traffic load inside the VoD system increase as directly proportion to the traffic arrival rate for $\lambda_i$ = 1, 1.5, 2, 2.5, 3, 3.5, 4, 4.5, 5, 5.5, 6.0…

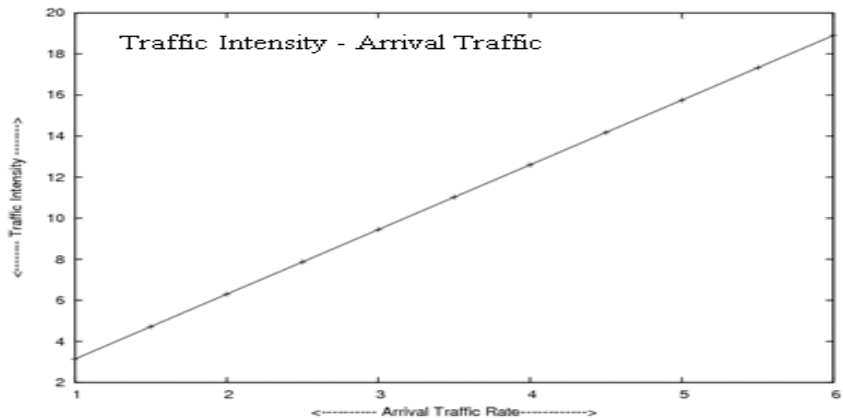

Figure 12

## 5. Concluding Remarks

In this work I have presented the performance of the distributed video on demand system with respect to the policy based traffic handle for the different topological networks architecture. In this work I have consider the different sized of the cluster populations. The clients of different population size submit their request with variable rate to the distributed video on demand system. The simulation result shows, the impact of policy based traffic handle on the performance of the video on demand system. The simulation result clearly reflect the efficiently used of policy based traffic handle remarkable reduced the traffic load inside the video on demand system. The mechanism also enhanced the overall system performance for the distributed video on demand system. In further multirate multisession request from the cluster of population be done with respect to the load balancing at the servers of the distributed VoD system. The algorithm to be develop for efficiently implement the multirate multisession in the distributed video on demand system.

**Author**

Soumen Kanrar received the M.Tech. Degree in Computer Science from Indian Institute of Technology Kharagpur India in 2000. Advanced Computer Programming RCC Calcutta India 1998. and MS degree in Applied Mathematics from Jadavpur University India in 1996. Currently he is working as research stuff at the Vehere Interactive Pvt Ltd - India. He has worked as the faculty at the department of Computer Science & Engineering DIATM, BCREC, AEC –INDIA, & King Saud University at Saudi Arabia.

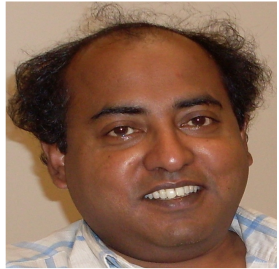